\documentstyle[11pt,epsfig]{article}
\evensidemargin\oddsidemargin
\begin{document}
\pagestyle{plain}

\title{\bf ZDC Effective Cross Section for Run 12 Uranium-Uranium Collisions in RHIC}
\author{A. Drees}
\date{\today} 
\maketitle
\section{Introduction}
An accurate calibration of the luminosity measurement of the 
2012 Uranium-Uranium RHIC run at 96  GeV per beam is of the greatest
importance in order to measure the total uranium-uranium cross section with a
reasonably small error bar. During the
run, which 
lasted from April 20th to May 15th 2012, three vernier scans per experiment were performed. 
Beam intensities of up to 3.4 10$^{10}$ Uranium ions in one ring were
successfully accelerated to flattop at $\gamma = 103.48$ corresponding
to 96 GeV/beam. The desired model
$\beta^*$ value was 0.7 m in the two low beta Interaction
Points IP6 and IP8. With these beam parameters 
interaction rates of up to 15 kHz were achieved. This
note presents the data associated with the vernier scans, and
discusses the results and systematic effects.

\section{Vernier Scan Technique}
During a vernier scan, also known as a Van der Meer scan~\cite{VDM},
the transverse size and shape of the beam overlap region is 
measured by recording the interaction rate as a function of
the transverse beam separation. A Gauss-fit of the measured interaction
rate as a function of the separation allows to determine the
effective beam size as well as the maximum achievable collision
rate  and thus the effective
cross section of the detector in use. Typically 
additional effects such as the presence of debunched beam, a possible crossing angle or the 
hourglass effect~\cite{hg} require correction factors to be
applied to the results. In the case of the stochastically cooled UU
beams~\cite{sc}, a simple Gauss-fit was 
not sufficient to describe the shape of the overlap region and a
2-Gauss fit had to be used to describe the data. The transverse beam
size was derived by using a weighed average of the widths of two
Gauss-functions.  
Table~\ref{tab:scanlist} lists all available vernier scans, the time
the data was taken, the IP, the fill pattern and which beam was moved
during the scan.
\begin{table}[h!]
\renewcommand{\arraystretch}{1.3}
\begin{center}
\begin{tabular}{|l|c|c|c|c|c|c|} \hline
{\bf fill} &{\bf date } & {\bf time} & {\bf ring } & {\bf IP } &{\bf fill pattern } &{\bf $n_{coll}$ } \\ \hline \hline
16783    & 04/25 & 19:31 - 19:49 & B & IP6 & 111x111 & 102 \\ \hline
16783    & 04/25 & 19:54 - 20:13 & B & IP8 & 111x111 & 111 \\ \hline
16842    & 05/09 & 12:34 - 12:53 & Y & IP6 & 111x111 & 102 \\ \hline
16842    & 05/09 & 12:55 - 13:11 & Y & IP8 & 111x111 & 111 \\ \hline
16857    & 05/14 & 15:21 - 15:34 & Y & IP6 & 111x111 & 102 \\ \hline
16857    & 05/14 & 16:08 - 16:20 & Y & IP8 & 111x111 & 111 \\ \hline
\end{tabular}
\caption{\label{tab:scanlist} List of fills with vernier scans
  during the 2012 UU run at 96 GeV/beam. }
\end{center}
\end{table}

\subsection{The Fit Function}
The shape of the overlap region, i.e. the collision rate $R_{coll}(x)$ as a function
of distance between the two beams, $x$, is mapped by the vernier scan
and can usually best be
described by a single Gauss-function:
\begin{equation}
R_{\rm \rm coll}(x) \; = \; R_{\rm Bkgd} \; + \; R_{\rm max} \; \times \;  \exp \left(  \frac
{-(x-x_0)^2}{2 \; \sigma_{\rm x}^{\rm VS}}  \right)
\label{eq:1G}
\end{equation} 
with the 4 free parameters: \\
$R_{\rm Bkgd}$: non-collision related background signal in the collision
rate (``offset'')\\
$R_{\rm max}$: maximum collision rate seen by the ZDC detector (corrected
for background) \\
$x_0$: location of the maximum \\
$\sigma_x$: width of the overlap region \\
\par
\begin{figure}[h!]
\begin{center}
\mbox{\epsfig{file=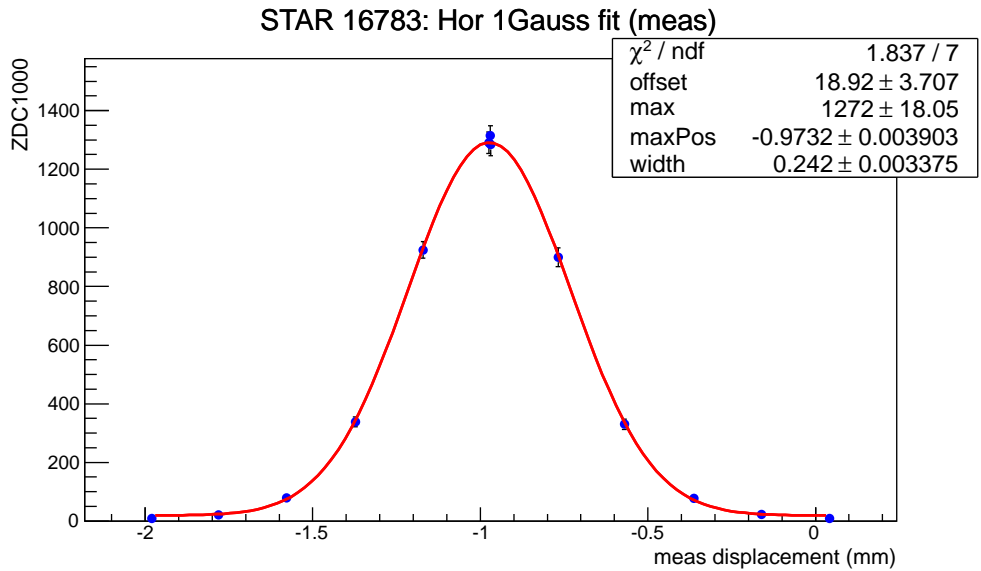,width=0.90\linewidth}}
\end{center}
\vspace*{-0.5cm}
\caption{\label{1Gfit} Horizontal data from a vernier scan in fill
  16783 at the STAR experiment fitted with a 1-Gauss 
  fit function. Stochastic cooling was not used. } 
\end{figure}

This approach is used when fitting data from a store with uncooled
Uranium beams (such as store 16783). Fig.~\ref{1Gfit} shows the
horizontal data
and 1-Gauss fit to the STAR ZDC data  during
the 16783 vernier scan.  
\par
However, in the case of stochastically cooled Uranium beams the single Gauss
approach does no longer fit the data 
and a double Gauss-function is chosen:
\begin{equation}
R_{\rm coll}(x) \; = \; R_{\rm Bkgd} \; + \; R_{\rm max,1} \; \times \;  \exp \left(  \frac
{-(x-x_0)^2}{2 \; \sigma_{\rm x,1}^{\rm VS}}  \right) \; + \; \; R_{\rm max,2} \;
\times \;  \exp \left(  \frac {-(x-x_0)^2}{2 \; \sigma_{\rm x,2}^{\rm VS}}  \right)
\label{eq:2G}
\end{equation} 
The double Gauss function has 6 free parameters: \\
$R_{\rm max,1}$: maximum collision rate of the core region \\
$x_0$: location of both the maxima \\
$\sigma_{\rm x,1}$: width of the core overlap region \\
$R_{\rm max,2}$: maximum of the tail distribution \\
$\sigma_{\rm x,2}$: width of the tail region \\
$R_{\rm Bkgd}$: non-collisional background rate (``constant'') \\
\par
The double Gauss approach was first necessary during the pp run at 250
GeV in Run-9~\cite{2Gauss}.  In this analysis it is assumed that both
Gauss distributions, core and tail,  are centered around the same 
location $x_0$. Data from stores 16842 and 16857 were fitted with the double Gauss. 
\begin{figure}[h!]
\begin{center}
\mbox{\epsfig{file=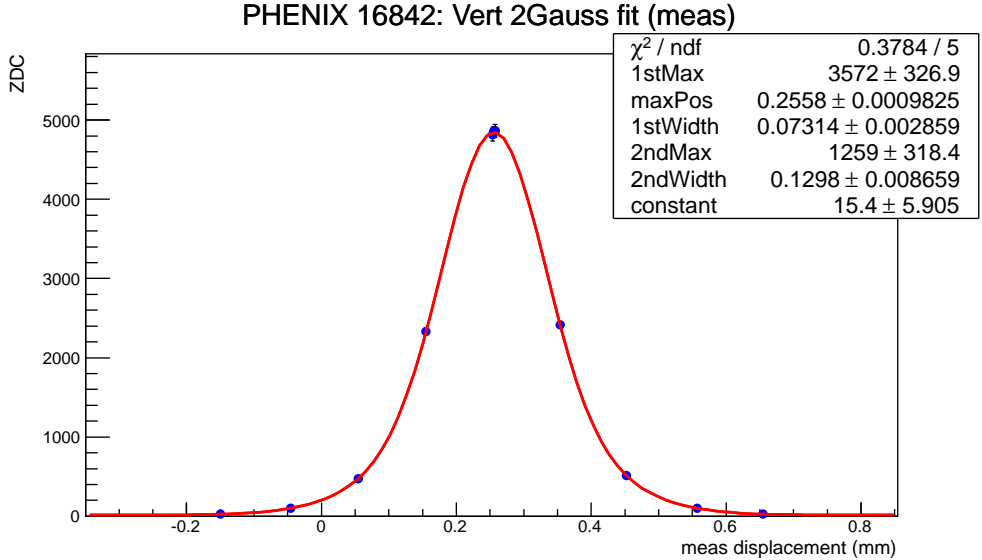,width=0.90\linewidth}}
\end{center}
\vspace*{-0.5cm}
\caption{\label{2Gfit} Vertical data from a vernier scan in fill
  16842 at the PHENIX experiment fitted with a 2-Gauss 
  fit function. Stochastic cooling was used during this store. } 
\end{figure}
Fig.~\ref{2Gfit} shows the vertical PHENIX ZDC data together with a
2-Gauss fit  during
the 16842 vernier scan. The beams were cooled during this store. 

\section{The Effective Cross Section}
The effective detector cross section $ \sigma_{\rm ZDC}^{\rm eff} $ of a
detector, in this case the ZDCs, can be
determined by the beam current, the collision rate and the overlap 
region $\sigma_{\rm x,y}^{\rm VS}$ of the two beams. The maximum collision
rate and the width of the overlap region are derived from the fits to
the vernier scan data.   
\begin{equation}
\label{eq:xsec}
\sigma_{\rm ZDC}^{\rm eff} = \frac{ R_{\rm max} \; 2 \pi \; n_B \; n_Y \; \sigma_{\rm x}^{\rm VS} \; \sigma_y^{\rm VS} }
          { n_{\rm coll} \; f_{\rm rev} \; N_B \; N_Y }
\end{equation} 
where: \\
$R_{\rm max} = $ maximum collision rate seen by the ZDC detector (corrected for background) \\
$n_B, \; n_Y = $ number of blue and yellow bunches respectively \\
$\sigma_{\rm x,y}^{\rm VS} = $ RMS beam-overlap size, derived from the fit to the vernier scan data\\
$n_{\rm coll} = $ number of colliding bunch pairs in the IP where the ZDC detector is located\\
$f_{\rm rev} = $ revolution frequency, approx. 78.4 kHz\\
$N_{\rm B,Y} = $ total number of ions in the blue and yellow rings, from WCM\\
\par
Fig.~\ref{2Gfit} shows one example of a vernier scan, here in
the vertical plane in IP8, moving the yellow beam. It
depicts the ZDC coincidence rate as a function of the measured
distance of the two beams. The data is fitted with a double
Gauss function according to Eq.~\ref{eq:2G}. The beam overlap size $\sigma^{\rm VS}$ in
Eq.~\ref{eq:xsec} is given by the combination of the two widths
from the double Gauss fit. The combined width adds the two individual widths according
to their amplitudes:
\begin{equation}
\label{eq:widthcomb}
\sigma_{\rm x,y}^{\rm comb} = \sigma_1^{\rm x,y} \; \frac{R_{\rm max,1}}{R_{\rm max,1} +
  R_{\rm max,2}} + \sigma_2^{\rm x,y} \; \frac{R_{\rm max,2}}{R_{\rm max,1} + R_{\rm max,2}}
\end{equation} 
``x'' and ``y'' refer to the two planes, while ``1'' and ``2'' refer
to the core (1) and tail (2) part of the distribution
respectively. 
\par
Several
corrections apply to the cross section  measurement
using the vernier scan method. They potentially affect the collision
rate (crossing angles, accidental coincidences, background), the beam
current measurement (fill pattern and 
debunched beam) and the measurement of the width (beam-beam, beam
separation measured by BPMs). The various effects are discussed below with the
exception of accidental coincidences (see~\cite{tnpp} for details) and
backgrounds. Non-collisional backgrounds are accounted for in
the fit-functions and systematic errors due to accidental coincidences are negligible with
collision rates of just a few 10 kHz. Regardless, the 
collision rates in the following are corrected for accidental coincidences.

\section{Corrections and Systematic Errors}

\subsection{Beam Current Measurement}
Due to Intra Beam scattering (IBS~\cite{ibs}) with heavy ions, a
standard heavy ion store of a typical length of about 6 to 7 hours is
subject to significant debunching of the beam. ``Standard'' in this context refers
to uncooled beams. However, during the most part of the
UU run in 2012 the heavy ions were not only stochastically cooled
in the longitudinal but also the two transverse
planes~\cite{sc}. If debunched beam, or beam which does not
contribute to collisions, is present in RHIC uranium-uranium stores,
this analysis cannot rely 
on the measurement of the total circulating beam current. Thus it is
mandatory to find proof for the presence or absence of debunched beam
before we decide on which device to rely on for beam current
measurement:  the DCCT~\cite{dcct} or the Wall Current Monitor (WCM~\cite{dcct}). While
the DCCT measures the total current circulating in RHIC, the WCM is
sensitive to only the bunched portion of it.
The differences between the DCCT and WCM
include possible true {\bf debunched beam}, a store to store variation of the
{\bf WCM calibration} to the DCCT and changes in the WCM readings due to
{\bf gain changes in the scopes}. These features are discussed below. 

\subsubsection{\label{calib}Calibrating the WCM}
Two scopes are used as the data acquisition system for the
WCMs. While an excess of bunched beam over total circulating beam has
to be attributed to a calibration issue, it is of course possible that the amount of bunched beam is
less than the total circulating beam. However, this does not apply to the period of
time when 
RHIC is ramping when accelerated beam has to be bunched (or get lost). Therefore the WCM can be
calibrated with the DCCT at the end of the ramp. For this analysis,
ev-flattop was chosen to signify the end of the
ramp. Tab.~\ref{tab:wcmcalib} shows the difference DCCT-WCM for 
the three vernier scan stores at ``flattop''.  
\begin{table}[h!]
\renewcommand{\arraystretch}{1.0}
\begin{center}
\begin{tabular}{|l|c|c|c|c|c|c|} \hline
{\bf fill} &{\bf Blue [ions x $10^{9}$] } &{\bf Blue [\%] } & {\bf Yellow [ions x $10^{9}$ ]} & {\bf Yellow [\%]} \\ \hline
16783    & -0.33 & -1.25 & -0.4 & -1.54 \\ \hline
16842    & -0.39 & -1.25 & -0.46 & -1.53 \\ \hline
16857    & -0.38 & -1.17 & +0.22 & +0.64 \\ \hline
\end{tabular}
\caption{\label{tab:wcmcalib}Blue and yellow WCM calibration for the
  three vernier scan stores in units of ions $ x 10^{9}$ and percent of
  total beam respectively. }
\end{center}
\end{table}
A negative sign signals that the WCM reading is above the DCCT reading
at the time the ramp reaches ``flattop''. Looking over a random
selection of other UU stores revealed large store to store variations
and a needed correction of the WCM readings of up  
to 2\%. However, the values found in the first two vernier scan
stores appear to be rather typical. Due to the availability of the
DCCT (with a factory listed absolute accuracy of 0.2\%) the varying calibration
of the WCM could be corrected for with an associated systematic error
of 2 x 0.2\%.
\par
\begin{figure}[h!]
\begin{center}
\mbox{\epsfig{file=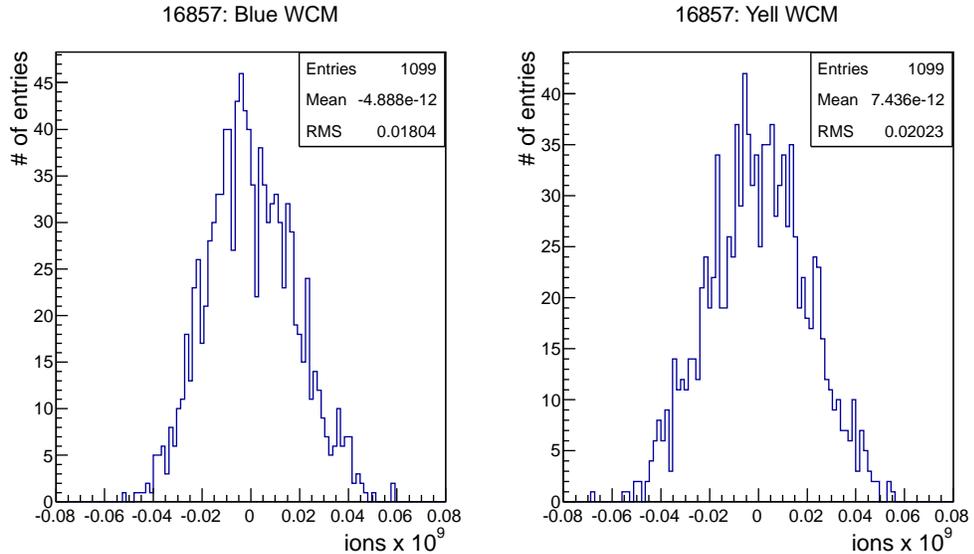,width=0.90\linewidth}}
\end{center}
\caption{\label{fig:wcmscatter}Point-to-point measurement scatter as
  seen by the blue (left) and yellow (right) WCM scope for the time
  period of about 1 h in store 16857. }
\end{figure}

 
In order to estimate the point-to-point measurement accuracy
or scatter of the WCM readings, a data set
from store 16857 corresponding
to about 1 h of data taking was selected. During this one hour neither the blue nor the
yellow scope did undergo a gain change and the beam current dropped
linearly by approximately 10\% in this time period. The linear drop
was fitted and the fit subtracted from the data to compensate for
ordinary beam loss in the data sample. The resulting two 
distributions are shown in Fig.~\ref{fig:wcmscatter}. 
The RMS of the scatter, 0.02 ions x $10^9$, is identical for the two
scopes corresponding to a 0.2\% effect per ring given a minimum beam
current of 12 x $10^9$ ions at the end of some UU stores. Thus a combined systematic 
error of 2 x 0.2\% is assigned to cover the scatter in the WCM
point-to-point measurements. 

\subsubsection{Debunched Beam}
\begin{figure}[h!]
\begin{center}
\mbox{\epsfig{file=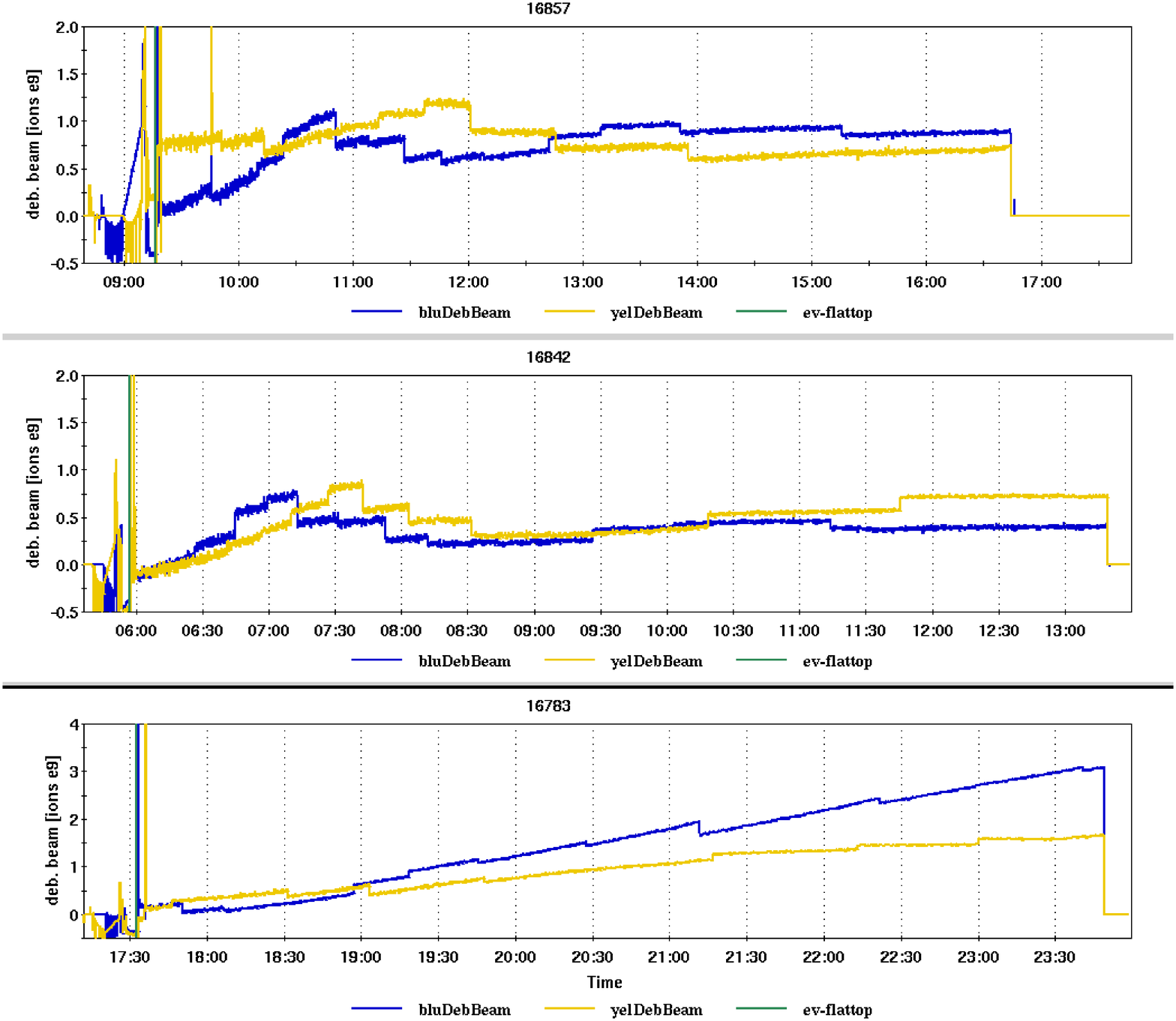,width=0.90\linewidth}}
\end{center}
\caption{\label{debBeam} Difference of the DCCT and WCM readings as a
  function of time for the three stores with vernier scans. A positive
  reading or 
slope indicates an increase of debunched beam in RHIC, equivalent to a
loss of bunched beam.}
\end{figure}

One indicator for the presence of true debunched beam is an apparent
difference between the DCCT and WCM, especially if this difference is
not constant throughout a store as would be the case with a simple
calibration offset. Fig.~\ref{debBeam} contains three graphs 
showing the difference between the two available devices, DCCT and WCM,
in the two rings for the three vernier scan stores. Units are number
of \mbox{ions x $10^{9}$}. The bottom 
graph, showing the store with the first vernier scan, depicts a steady
and linear increase of the amount of debunched beam due to the absence
of cooling. Clearly, the WCM has to be used to analyse data from this
scan. The center and top graph, however, showing the two other stores
with vernier scans, depict much less of an increase but nevertheless a steady
upward slope within the first 1 to 2 hours of the store. The
readings are stable after that (neglecting the effect of gain changes
for this discussion, see section~\ref{gains}). The time frame of 1 to
2 hours corresponds to about the time it 
took for the bunch length to reach the initial value again after a
period of shortening and lengthening due to longitudinal cooling. 
\par
In addition to evidence of some ongoing debunching during the first hours of a
store, there appears to be some loss of bunched beam associated with
the rebucketing exercise. Rebucketing is performed as soon as the ramp
is finished, within less than 100 sec after reaching flattop. The loss
of beam at rebucketing in the three vernier scan stores is summarized in
Tab.~\ref{tab:rebucket}. According to the WCM, some beam is lost in
all three cases.  Note that going through rebucketing is always
accompanied by a gain change in the WCM scopes. Gain changes in the scopes
are known to cause distinct changes in the WCM readings that are
solely due to the change in the scope gain and not due to a true
change of bunched beam. The effect of gain changes is discussed in
more detail in section~\ref{gains} below. Such distinct changes in the
WCM readings (increase as well as decrease) can
easily be seen in Fig.~\ref{debBeam} above for all three stores.
\par 
\begin{table}[h!]
\renewcommand{\arraystretch}{1.0}
\begin{center}
\begin{tabular}{|l|c|c|c|c|c|c|} \hline
{\bf fill} &{\bf Blue [ions x $10^{9}$] } &{\bf Blue [\%] } & {\bf Yellow [ions x $10^{9}$ ]} & {\bf Yellow [\%]} \\ \hline
16783    & -0.530 & -1.97 & -0.306 & -1.19 \\ \hline
16842    & -0.344 & -1.15 & -0.372 & -1.23 \\ \hline
16857    & -0.405 & -1.17 & -0.484 & -1.45 \\ \hline
\end{tabular}
\caption{\label{tab:rebucket}Loss of bunched beam at rebucketing in
  units of ions x $10^{9}$ and percent of 
  total beam as measured by the WCM. }
\end{center}
\end{table}
\begin{figure}[h!]
\begin{center}
\mbox{\epsfig{file=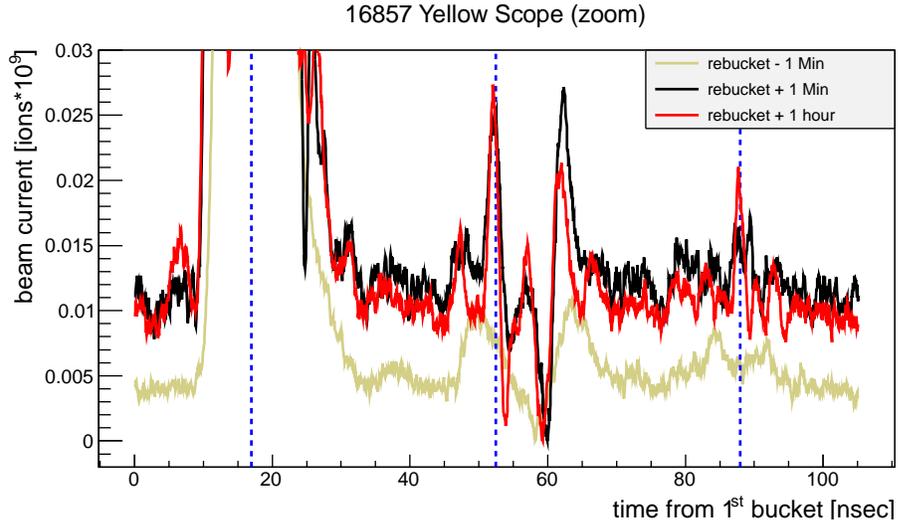,width=0.90\linewidth}}
\end{center}
\caption{\label{fig:yscope}Three traces from the yellow WCM scope, each combining all 120
  buckets in RHIC, taken at various times before and after
  rebucketing. Vertical scale is zoomed for better visibility. }
\end{figure}
The amount of beam loss at rebucketing is significant in all three
cases in Tab.~\ref{tab:rebucket}. If it is
real, the beam could be visible in the WCM scope. Fig.~\ref{fig:yscope}
shows three scope traces from the yellow scope in store 16857. Each
trace corresponds to an average of all 120 sets of consecutive 3
buckets~\cite{mb} in RHIC where only the first of the 3 buckets is expected to
be filled with beam. The vertical dashed blue lines correspond to the
centers of the 3 buckets. 
The first trace (tan colored), taken about 1 minute before
rebucketing, shows no apparent signal besides the main bunch in bucket
\#1 of the 3 shown buckets and some noise. The second trace (black) is taken 1
minute after rebucketing. Some beam appeared in the neighboring bucket
\#2  together with a pattern (centered around 60 nsec) that can perhaps be attributed to
ringing in the scope signal. The third trace (red) is taken 1 hour after
rebucketing. Enough time passed for cooling to take effect and trap
beam in any of the 3 buckets. A small but centered signal appeared in
bucket \# 3. This beam, while not debunched in the literal sense,
will not contribute to collisions. 
\par
Due to the incontrovertible presence of some, however small, amounts
of debunched beam in UU stores, the WCM will have to be used for this
analysis in lieu of the more accurate DCCT. 
    
\subsubsection{\label{gains}Gain Changes}
As stated above, the rebucketing procedure is always accompanied by a
gain change in the WCM scopes. Since the WCM has to be used it is of
interest how much of the rebucketing loss is real rather than a gain
change artifact. In order to estimate which portion of the apparent
beam loss is debunched beam and how much is due to a gain change, a
total of 36 UU stores were analyzed. In addition to the rebucketing
gain change there are approximately 10 more asynchronous gain changes
later during the store. Most of them cause the WCM reading to change
distinctly. The WCM changes have both signs. For this analysis 7 stores were randomly
selected in addition to the 3 stores with vernier scans yielding 108
blue scope gain changes at store and 93 yellow scope gain changes at store. The gain
changes at rebucketing are not part of this set. Blue and yellow
datasets were combined since both reveal the same general behavior.  
\begin{figure}[h!]
\begin{center}
\mbox{\epsfig{file=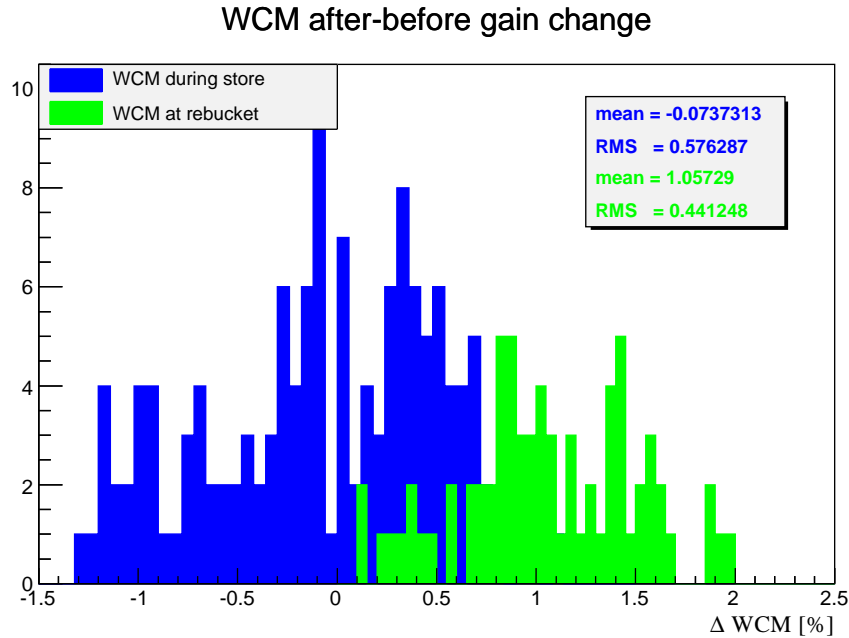,width=0.90\linewidth}}
\end{center}
\caption{\label{fig:gains}Histogram of two datasets containing scope
  gain changes (blue) and rebucketing changes (green). Blue and yellow
data are combined.}
\end{figure}
Fig.~\ref{fig:gains} contains histograms of both datasets, at store
and at rebucketing. The
blue shaded histogram shows the number of entries with distinct
changes in the WCM readings 
caused by a gain change of the scope while at store. The amount of
change is given in percent of beam current at the time and is
calculated as the difference of the reading after and before the scope
changed its gain. The mean and RMS value of this distribution are
printed in blue inside the graph. The mean is consistent with 0 with
an RMS of 0.58\%. The green shaded histogram shows the number of
entries with the quoted amount of change in the WCM readback that is
associated with rebucketing. 
\begin{figure}[h!]
\begin{center}
\mbox{\epsfig{file=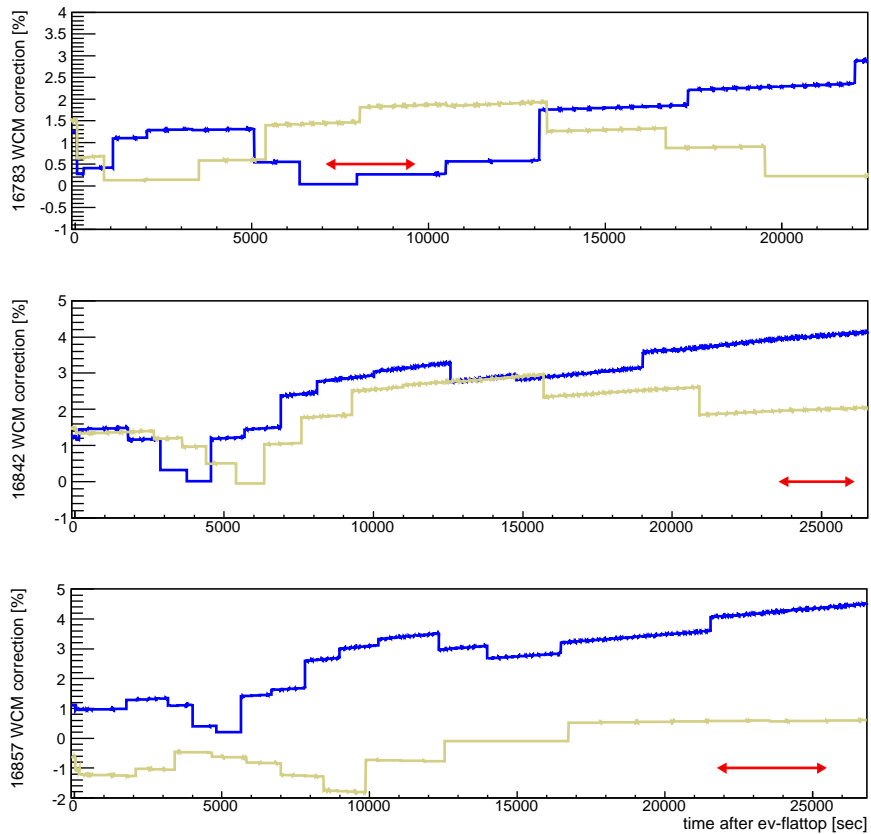,width=0.90\linewidth}}
\end{center}
\caption{\label{fig:corrwcm} Size of the correction to the WCM
  readings for the blue and yellow rings at store as a function of time after
  reaching ``flattop'' for the three vernier scan stores.}
\end{figure}
The mean is
shifted from 0 to +1\%, with an RMS of 0.4\%. This distribution is
clearly not consistent with the other. While the $\Delta$WCM due to store gain
changes scatters around 0, this distribution shows a mean value that is
not consistent with 0 but with an average bunched beam loss of
1\%. This signifies a true loss of bunched beam at rebucketing of an
average of 1\% of the beam current which is superimposed by a WCM
reading shift triggered by the scope gain change. The extra
fake shift causes the measured beam loss at rebucketing to be decreased or
increased, depending on the random nature of the shifts caused by gain
changes. Thus, for this analysis, the bunched beam current at rebucketing is
reduced by 1\% for each individual vernier scan store with an associated
systematic error of 0.4\% per beam totaling 0.8\%. 
\par
What is left to do is a correction for all following gain changes
during the store based on the assumption that none of the associated
WCM shifts are real. The first step is a correction of the
WCM reading based on the calibration with the DCCT at the end of the
ramp (see section~\ref{calib} above). This is followed by the 1\%
correction at rebucketing as described above. The effect of each following gain
change is erased assuming that the calibrated reading from before the
gain change is accurate. Fig.~\ref{fig:corrwcm} shows the accumulating
corrections in percent for the three vernier scan stores. 
The time ranges during which the vernier scans happened are indicated
by the red arrows. The corrections to the blue and yellow WCM readings
are shown as a blue and yellow line respectively. At the time of the
vernier scan the corrections range from practically 0\% to up to 4\%
(blue in 16842 and 16857). 
\begin{figure}[h!]
\begin{center}
\mbox{\epsfig{file=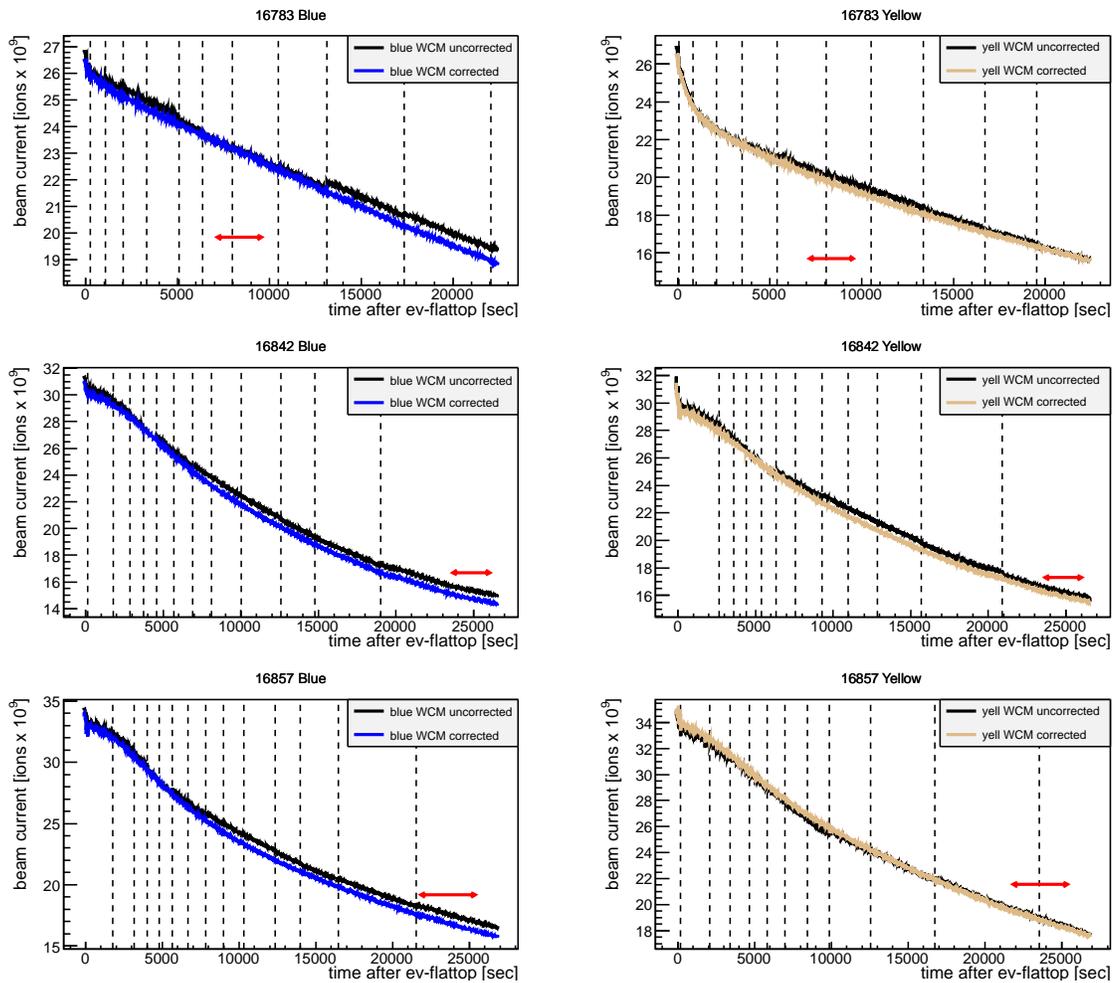,width=1.05\linewidth}}
\end{center}
\caption{\label{fig:plotwcm} Corrected and uncorrected WCM readings as
a function of time after ``flattop'' for the three vernier scan
stores. The time of gain changes is indicated by vertical dashed lines.}
\end{figure}
Fig.~\ref{fig:plotwcm} shows the corrected
and uncorrected beam current as measured by the WCM with and without
corrections (as shown in Fig.~\ref{fig:corrwcm}) for the blue and yellow
rings. Blue is shown on the left 
while yellow is shown on the right. All three stores with vernier
scans are included, the times of the vernier scans are again added as
red double arrows. The associated systematic error is based on the RMS
of the
blue shaded distribution shown in 
Fig.\ref{fig:gains} and totals 1.2\% for the two beams together.

\subsubsection{\label{sec:fillpatt}Fill Pattern}
\begin{figure}[h!]
\begin{center}
\mbox{\epsfig{file=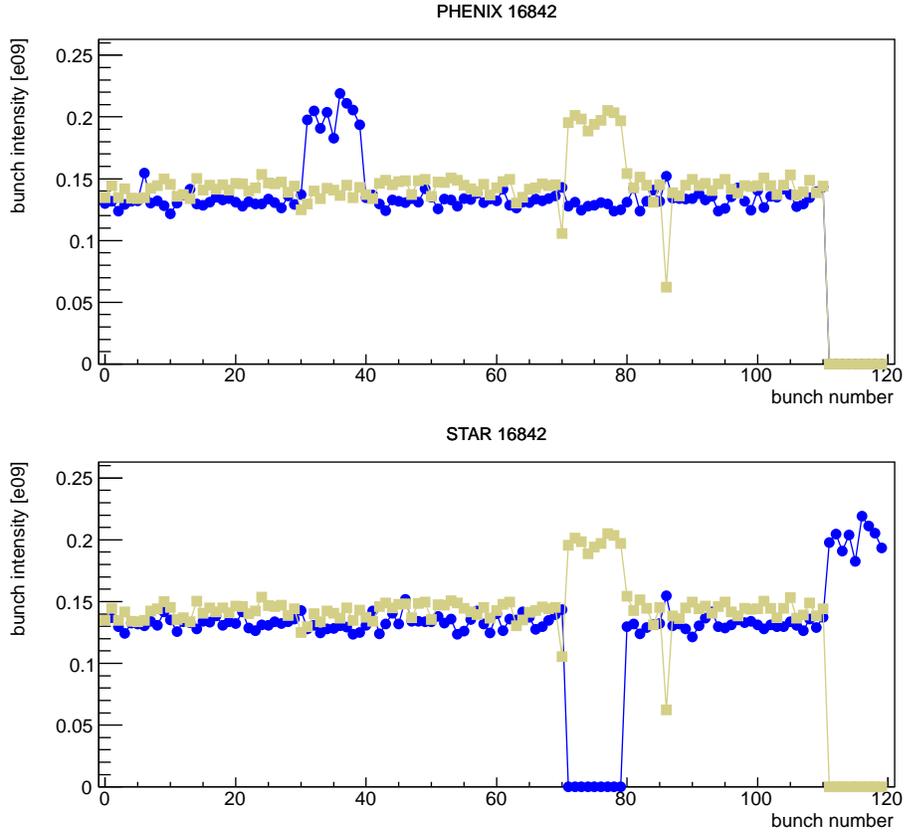,width=0.90\linewidth}}
\end{center}
\caption{\label{fillpatt}Bunch intensity as a function of yellow bunch
  number for the two beams according to colliding pairs in PHENIX
  (top) and STAR (bottom). }
\end{figure}

In Eq.~\ref{eq:xsec} an average beam intensity per bunch is
used independent of the actual bunch pairs colliding at a given
IP. Generally this is a very good approximation but the true value
can deviate in case of very unevenly filled bunches. This is indeed
the case for stores 16842 and 16857 in particular and cooled stores in 
general. Fig.~\ref{fillpatt} shows store 16842 as an example. The top graph    
shows the bunch intensity for both beams as a function of the yellow
bunch number. In the PHENIX IP blue bunch \#1 collides with yellow bunch
\#1 and the abort gaps line up.  The bottom graph also shows the bunch intensity
for both beams as a function of the yellow bunch number. In this
graph, however, the blue bunch train is shifted by 60 degrees such that
blue bunch \#41 collides with yellow bunch \#1. The abort gaps do not
line up causing some yellow and some blue bunches to collide with the
gap in the other beam. 
Due to the different levels of burn-of the bunches with only one
collision clearly show a signficantly 
higher bunch intensity than the other quite evenly filled
bunches. In this snapshot the intensity of one-collision bunches exceeds the
one of bunches with two collisions by 30\%. This causes a significant
correction to the term $(n_B \; n_Y )/( n_{\rm coll} \; N_B \; N_Y )$
in Eq.~\ref{eq:xsec} for IP6 in the two stores with cooled 
beams. In this analysis, average bunch currents and thus cross section and
instantaneous luminosity measurements are
corrected for this effect. Correction factors vary between 0.2\%
(IP8, 16783) up to 6.6\% (IP6, 16842). No additional systematic uncertainty is
assigned to this effect since it can be measured and corrected for.

\subsection{Beam Position Monitors}
Beam position monitors enter a discussion of vernier scan uncertainties in two
ways: 
\begin{itemize}
\item How accurate is the determination of the distance between the
two beams during the vernier scan? 
\item Is there a crossing angle at the IPs reducing
  the maximum achievable luminosity?
\end{itemize}
For the first question only relative position measurements matter, i.e. how
well the BPMs are capable to measure a deliberate beam displacement of
a few hundred microns. The relative accuracy could be very good even
at the presence of a large absolute position error. The quality of the
relative measurement determines the accuracy of the width measurement
in the vernier scan. 
\par
For the second question the accuracy of the absolute position
measurements on either side of the IP matters. If significant true offsets of the
beam position on both or just one side of the IP are present a virtual
crossing angle could result as well as the opposite, i.e. a true crossing
angle could go unnoticed due to false position measurements. Both
types of measurement accuracies are discussed below.  
 
\subsubsection{Relative Measurements}
\begin{figure}[h!]
\begin{center}
\mbox{\epsfig{file=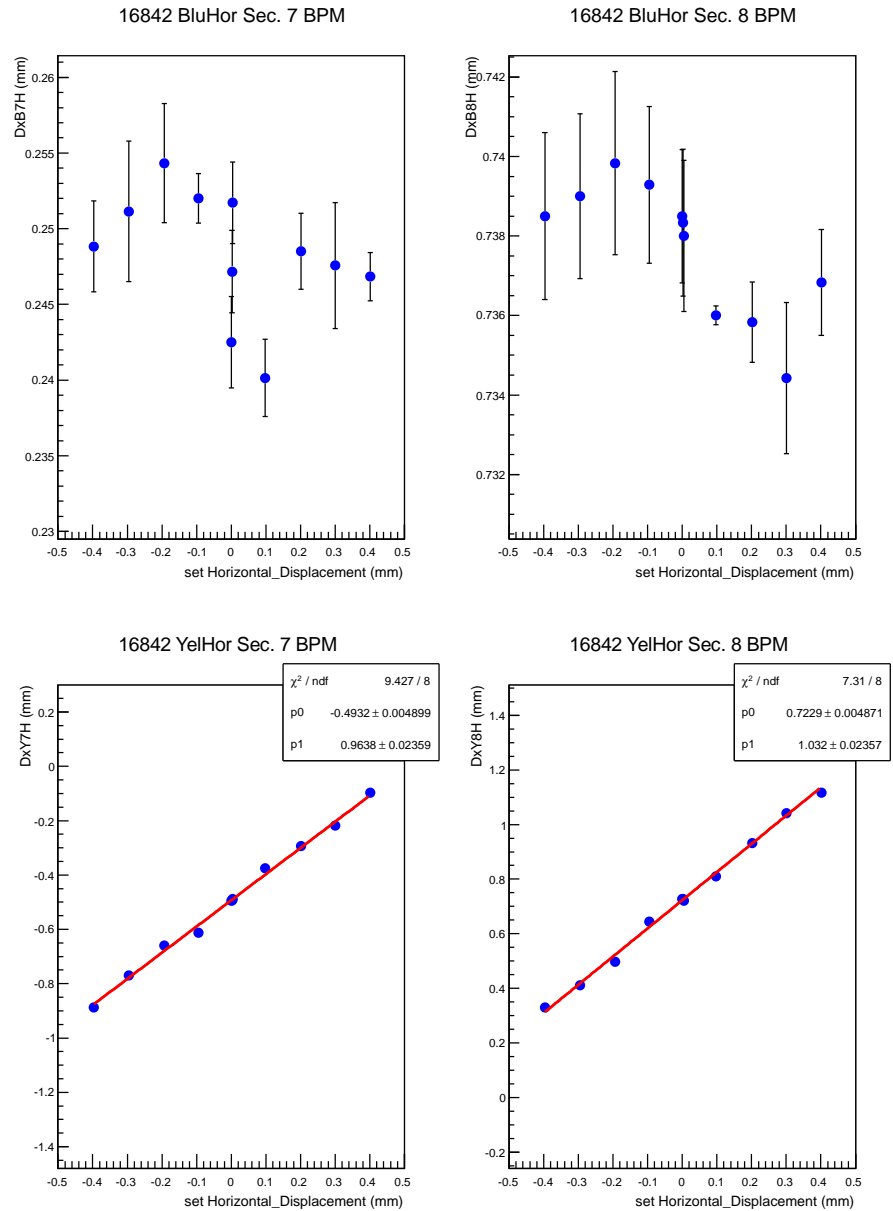,width=0.80\linewidth}}
\end{center}
\caption{\label{fig:HHIP8} Beam position measurements during the horizontal
part of the vernier scan in fill 16842 in IP8. All four horizontal
DX BPMs are presented. Top row shows blue BPMs, the bottom row
shows yellow BPMs. }
\end{figure}

During a vernier scan the {\bf wanted} beam position (i.e. the set
value for the scan request) as well as the {\bf measured} beam position are
recorded. Fig.~\ref{fig:HHIP8} shows a typical instance of such a data
set from the vernier scan in fill 16842. The Panel contains four
graphs, each showing the measured beam 
position for the blue beam (top) and the yellow beam (bottom) as a
function of the set value. The BPMs on the two sides of IP8 are shown
individually. 
In this particular example the yellow beam was scanned across the
blue. The consequent variation of the beam position in the blue ring
indicates a sufficient beam-beam interaction between the two beams to
cause a small effect. The
typical ``S''-shape is not present for the first scan when beams were
un-cooled and is less pronounced for the third. The resulting
variation even in the shown case is 
small, no larger 
than 15 $\mu$m peak to peak (and smaller than in the case of scanning
255 GeV protons~\cite{tnpp}). Nevertheless, it is taken into account
when the distance between the two beams is calculated. Therefore no
systematic error is dedicated to the beam-beam effect, instead it is
corrected for and covered with the relative position measurement error.  
\par
For the shown yellow data set the deviation of the fitted slope from ``1'' is
small, -3.5\% and +3\% for DxY7H and DxY8H respectively, thus resulting in an
agreement of the interpolated measured position at the IP to the set value of
better than 1\%. The interpolated measured position at the IP, 
${\rm Pos}_{\rm IP}$, is given by:
\begin{equation}
{\rm Pos}_{\rm IP} = \frac{1}{2} \; ({\rm BPM}_{\rm in} + {\rm BPM}_{\rm out}) ,  
\label{eq:posip}
\end{equation}
with ``in'' and ``out''
referring to the two sides  of the IP (in-coming and out-going).
Tab.~\ref{tab:slopes} summarizes the agreement of the measured
position at the IP with the set values 
for both IPs and all three scans. 
\begin{table}[h!]
\renewcommand{\arraystretch}{1.1}
\begin{center}
\begin{tabular}{|c|c|c|c|} \hline
    & 16783 & 16842 & 16857 \\ \hline
Hor IP6 & 1.006 & 0.99 & 0.99 \\
Ver IP6 & 0.99 & 1.0 & 1.00 \\ \hline
Hor IP8 & 1.006 & 0.998 & 1.00 \\ 
Ver IP8 & 1.013 & 1.02 & 1.016 \\ \hline
\end{tabular}
\caption{\label{tab:slopes} Slopes fitted to measured separation as a
  function of set separation. A slope of ``1'' corresponds to perfect
  agreement between the two. }
\end{center}
\end{table}
The slopes all scatter around ``1'' (i.e. perfect agreement) by about
$\pm$ 1\% and show no consistent affinity 
towards an over- or under-shoot. The measured scatter includes
beam-beam as well as potential hysteresis effects. Because of the excellent
agreement actual beam positions measurements were used in this
analysis (instead of the ``set'' values from the model) and a
systematic error of 1.0\% is assigned to the relative beam 
position measurement. No systematic error is dedicated to hysteresis specifically
since it is covered by the above quoted error of 1\%.

\subsubsection{Absolute Measurement}
\begin{figure}[h!]
\begin{center}
\mbox{\epsfig{file=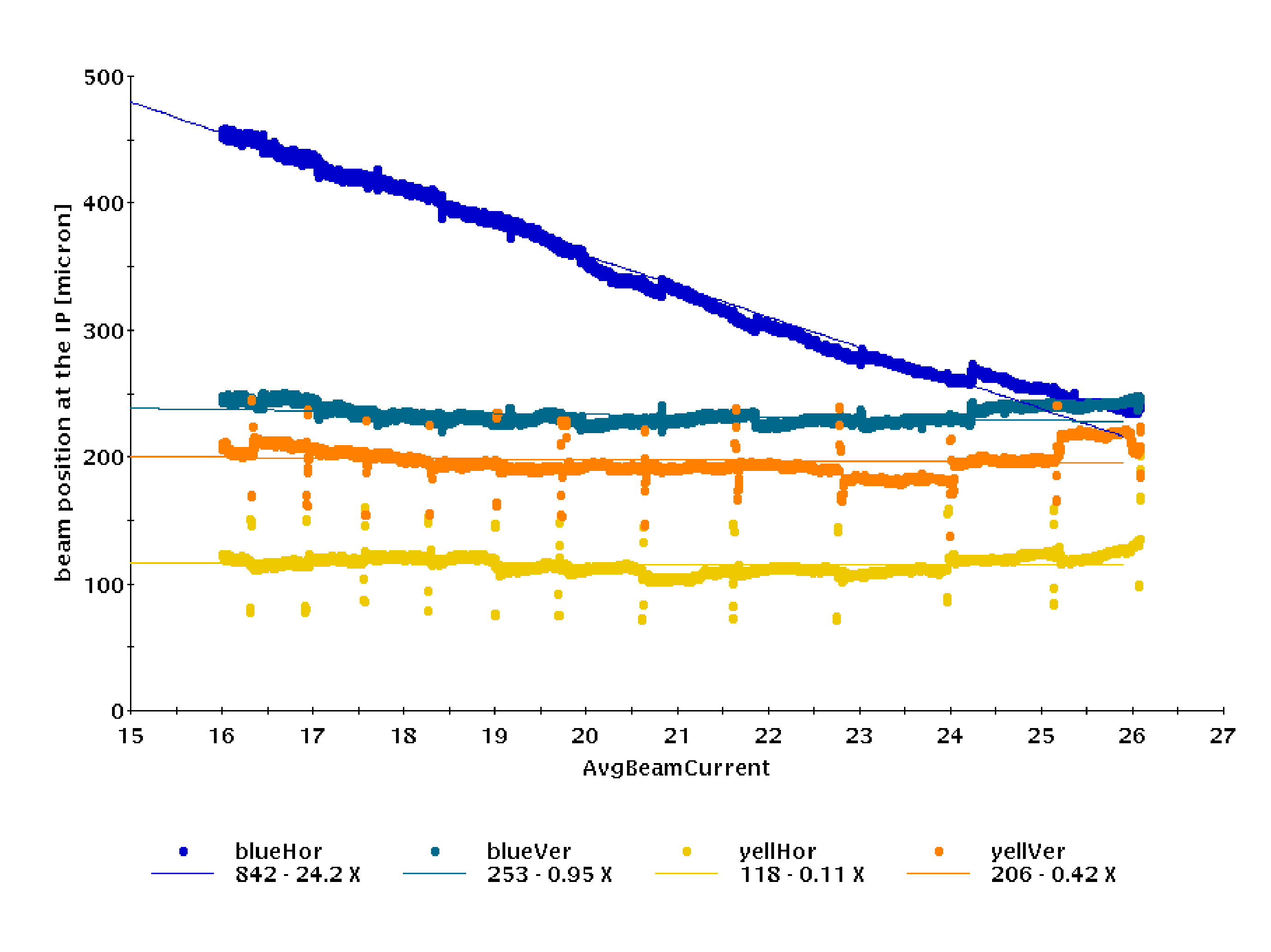,width=0.90\linewidth}}
\end{center}
\caption{\label{BBpm} Horizontal and vertical blue beam position at the PHENIX IP measured by
  the DX BPM as a function of total beam current.}
\end{figure}

The differences between the blue and yellow beam positions in the two
planes while beams are overlapping fully (i.e. after a successfull
luminosity optimization such as a vernier scan) indicate the range of
absolute beam position measurement errors since the expectation value
is ``0''. 
Fig.~\ref{BBpm} shows the measured blue beam
position at IP8 in the blue vertical plane (``blueVer''), the blue horizontal
plane (``blueHor''), the yellow horizontal plane (``yellHor'') and the
yellow vertical plane (``yellVer'') for fill 16850 as a function of
the combined blue and yellow beam
intensity (``AvgBeamCurrent''). 
The short-lived and regular excursions apparent in the yellow data are
a telltale of regular automatic luminosity optimizations. All four data sets are
fitted with a $1^{st}$ order polynomial. While both 
vertical data sets as well as the yellow horizontal data set are all
consistent with a constant, the blue horizontal 
dataset reveals an obvious dependency and increase of the measurement
with a decreasing beam current. In this particular case it appears as
if the beam
drifted by 220 microns during the store which would, for cooled beams,
correspond to more than 3 $\sigma$! With such a separation, the
instantaneous luminosity would have dropped to nothing provided that
the other beam's position doesn't drift as well. Clearly, however,  yellow
horizontal is as constant as the two vertical planes. In addition, 
the regular automatic orbit corrections in combination with luminosity
re-optimizations ought to keep the blue horizontal beam position
stable and luminosity maximized. Therefore the drift, that can be traced back to the blue 
horizontal DX BPM on the sector 7 side of PHENIX (DxB7H), has to be fake. 
\par
Most stores (approx. 75\%) in the
UU-run display this fake drift feature, with the yellow horizontal and vertical
position as stable as the blue vertical one in this picture and the
blue horizontal one drifting. Beam intensities in the UU run, with a typical
\begin{table}[h!]
\renewcommand{\arraystretch}{1.1}
\begin{center}
\begin{tabular}{|c|r|r|r|c|c|} \hline
    & 16783 & 16842 & 16857 & all stores & mean \\ \hline
Hor IP6 & 20 & 40 & -10 & -30 to 80 & 30 \\
Ver IP6 &-20 & -30 & -20 & -80 to 10 & -40 \\ \hline
Hor IP8 & 140 & 100 & 80 & 80 to 190 & 135 \\
Ver IP8 & 40 & 60 & 40 & 10 to 90 & 50 \\ \hline
\end{tabular}
\caption{\label{tab:diff} False beam separation measured by DX BPMs at the IPs after
  optimization in units of microns. }
\end{center}
\end{table}
bunch intensity at the beginning-of-store of 0.3 x $10^9$, were small
compared to Au-beams with up to 1.3 x $10^9$ ions/bunch in previous
years. Thus I assume that the fake drift is not only correlated with
but caused by the small and decreasing beam 
intensity although the measured drift slopes vary by more than a factor 2 from
store to store and
most DX BPMs seem not affected and not even DxB7H demonstrates it
in all stores. 
The described fake drift is present in all 3 vernier scan
stores. Therefore, to estimate the measured difference between the beams I will
use the position measurement with the highest beam intensity, i.e. the
beginning-of-store after the first optimization. Tab.~\ref{tab:diff}
summarizes the results. 
The mean measured beam separation varies between -40 micron and +135
micron depending on the plane and IP. At the same time we know the
expected true value with 
luminosity optimized beam positions is 0. The measurements scatter
from store to store
with $\pm$ 50 microns around this mean. Picking the worst case,
i.e. Hor IP8, yields 135 microns, an amount that 
could manifest itself on both sides of the IP. The corresponding
uncertainty in terms of a crossing angle yields 0.015 mrad. 
\par
\begin{figure}[h!]
\begin{center}
\mbox{\epsfig{file=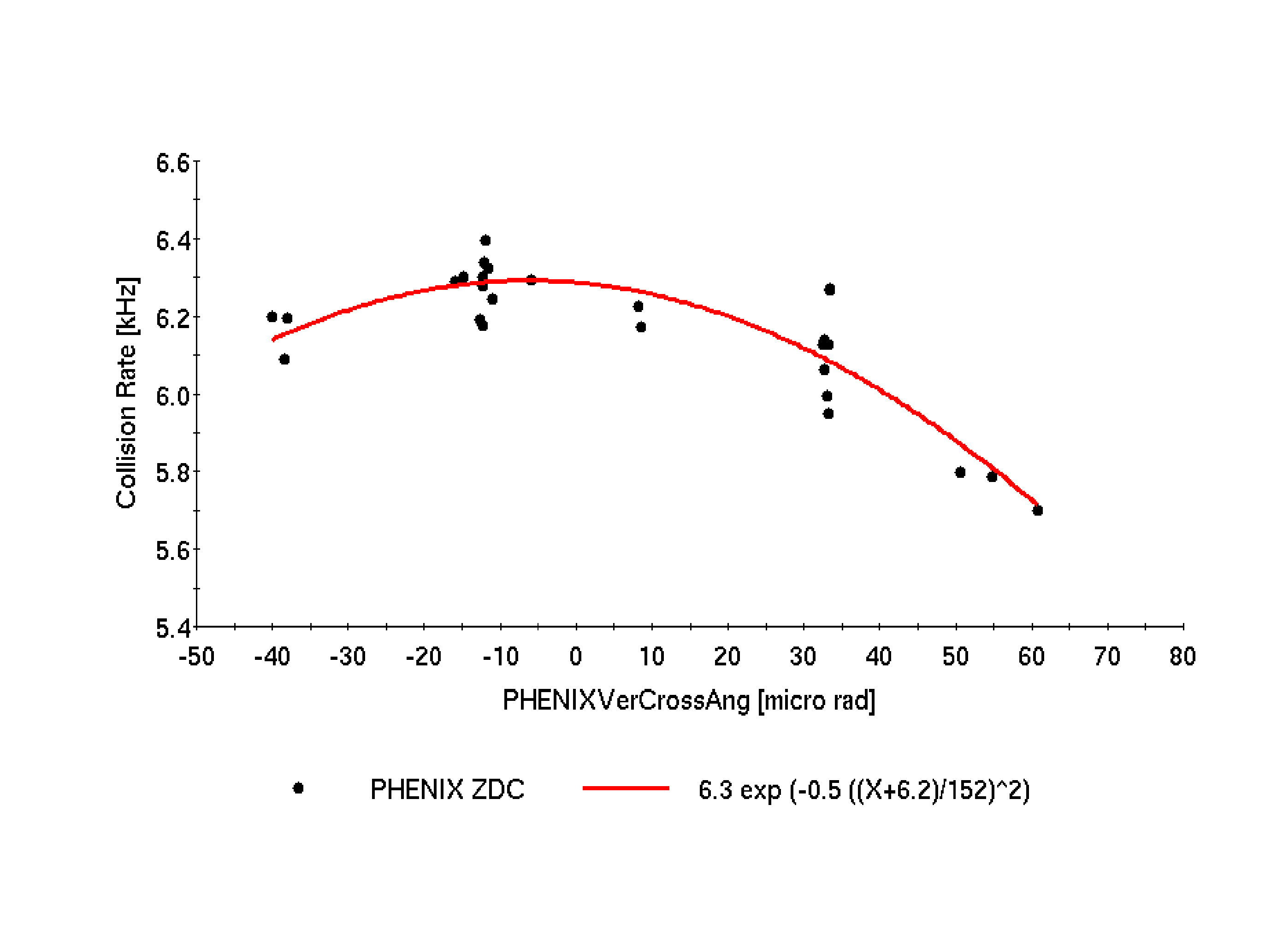,width=0.80\linewidth}}
\end{center}
\caption{\label{anglescan}Vertical crossing angle scan in store 16857
  in PHENIX. A Gauss-fit of the data is shown as well.}
\end{figure}
Fig.~\ref{anglescan} shows the PHENIX ZDC coincidence rate in units of
kHz as a function of the vertical crossing angle measured by the DX
BPMs in IP8. A Gauss-fit of the data is superimposed. In case of a
15 $\mu$rad crossing angle the achievable collision rate would be
reduced by 1\% from its maximum. In general, there could be a crossing
angle present in both planes independently, doubling the effect. Thus the uncertainty
due to unknown crossing angles totals 2\%. 

\section{\label{results}Results}
After applying all corrections the effective ZDC cross sections and instantaneous
luminosities for STAR and PHENIX are listed in
Tab.~\ref{tab:xsec}.
\begin{table}[h!]
\renewcommand{\arraystretch}{1.0}
\begin{center}
\begin{tabular}{|l|c|c||c|c|} \hline
 & \multicolumn{2}{|c||}{STAR} &  \multicolumn{2}{c|}{PHENIX} \\ \cline{2-5}
 fill & $\sigma_{\rm eff}^{\rm ZDC} [barn]$ & ${\cal L} [10^{\rm 25} cm^{\rm -2} s^{\rm -1}]$ &
 $\sigma_{\rm eff}^{\rm ZDC}$ [barn]& ${\cal L} [10^{\rm 25} cm^{\rm -2} s^{\rm -1}]$  \\ \hline
16783 & 15.70 $\pm$ 0.47 & 8.05 $\pm$ 0.24   & 16.09  $\pm$ 0.48 & 8.06 $\pm$ 0.24\\ 
16842 & 16.05 $\pm$ 0.48 & 28.00 $\pm$ 0.84  & 15.66 $\pm$  0.47 & 31.24 $\pm$ 0.62 \\
16857 & 15.78 $\pm$ 0.47 & 35.09 $\pm$ 1.05 & 15.65 $\pm$ 0.47 & 37.56 $\pm$
1.13 \\ \hline
\end{tabular}
\caption{\label{tab:xsec} Effective ZDC cross sections and instantaneous
  luminosities at the two IPs with statistical errors. }
\end{center}
\end{table}
\begin{figure}[h!]
\begin{center}
\mbox{\epsfig{file=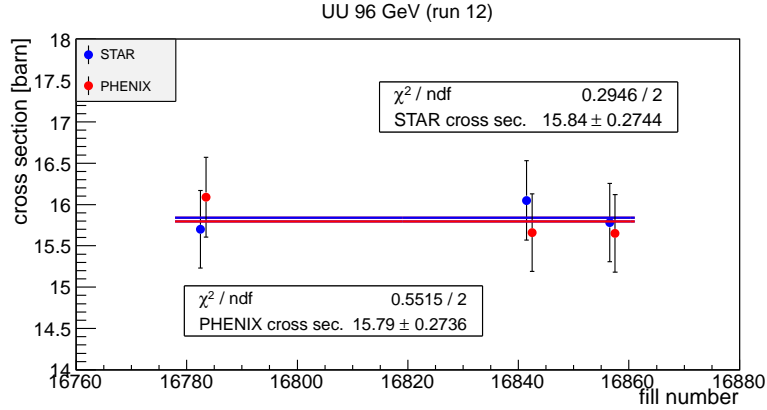,width=0.75\linewidth}}
\end{center}
\caption{\label{fig:xsec} Cross sections as measured by the 3 vernier
  scans as a function of the fill number. All corrections applied.}
\end{figure}
The three measurements per IP are all consistent with each other and
with statistical scatter.
The combined cross section measurements are shown in Fig.~\ref{fig:xsec} and yield
15.84 $\pm$ 0.27 b for STAR and 15.79 $\pm$ 0.27 b for PHENIX. The effective
cross sections of the two ZDCs are identical within the statistical errors. 
\par
Measured beam-overlap
sizes, another result from vernier scans, are summarized in
Tab.~\ref{tab:emit}. From the fit to the vernier scan data it 
is impossible to distinguish between the blue and yellow beam
therefore equal beamsizes were assumed when calculating the
emittance. The emittance values are hour-glass corrected and listed in
the right half of Tab.~\ref{tab:emit}. 
\begin{table}[h!]
\renewcommand{\arraystretch}{1.0}
\begin{center}
\begin{tabular}{|l|c|c|c||c|c|c|} \hline
 device & $\sigma^{\rm vs}$ 16783 & $\sigma^{\rm vs}$ 16842 &
 $\sigma^{\rm vs}$ 16857 & $\epsilon^{\rm RMS}$ 16783 & $\epsilon^{\rm
   RMS}$  16842 & $\epsilon^{\rm RMS}$ 16857 \\ \hline
Hor VS IP6 & 242 $\pm$ 7 & 95 $\pm$ 3 & 98 $\pm$ 3 & 2.9 $\pm$ 0.1 & 0.35 $\pm$ 0.01 & 0.38 $\pm$ 0.01 \\
Hor VS IP8 & 249 $\pm$ 7 & 94 $\pm$ 3 & 96 $\pm$ 3 & 3.0 $\pm$ 0.1 & 0.34 $\pm$ 0.01 & 0.36 $\pm$ 0.01 \\
Hor IPM & - & - & - & 2.2 $\pm$ 0.2 & 0.33 $\pm$ 0.03 & 0.3 $\pm$ 0.03 \\ \hline
Ver VS IP6 & 242 $\pm$ 7 & 90 $\pm$ 3 & 96 $\pm$ 3 & 2.9 $\pm$ 0.1 & 0.31 $\pm$ 0.01 & 0.37 $\pm$ 0.01 \\
Ver VS IP8 & 249 $\pm$ 7 & 88 $\pm$ 3 & 93 $\pm$ 3 & 3.0 $\pm$ 0.1 & 0.30 $\pm$ 0.01 & 0.34 $\pm$ 0.01 \\
Ver IPM & - & - & - & 2.8 $\pm$ 0.3 & 0.3 $\pm$ 0.03 & 0.35 $\pm$ 0.03 \\ \hline
\end{tabular}
\caption{\label{tab:emit}Measured size of the beam overlap region in units of
  $\mu$m and calculated RMS normalized emittances in units of mm
  mrad. Model beta functions were used. Supplied errors are
  statistical only. IPM emittances for the two rings were combined
  into one value assuming errors of 10\%.}
\end{center}
\end{table}
The values in this table were all calculated using a model $\beta^*$
value of 0.7 m for the two rings and both planes. Results from the
IPM~\cite{IPM} are added for comparison. The agreement is rather good
with the exception of the horizontal data from store 16783 and 16857 where
the IPM measurements are 25\% and 20\% below the results from the vernier
scan. This discrepancy cannot be explained by wrong beta-functions being
used since such beta-beat would not change from store to store. There
is no evidence of a deviation from the round beam assumption in the data, in fact
the horizontal and vertical beam sizes are identical within
statistical errors. There is also no evidence of the beam size being
larger in one IP versus the other indicating equal beta functions in
the two experimental IPs. 
\par
Nevertheless, the available data was
compared with measured beta-functions from~\cite{beta}. Measured and
model beta functions are summarized in Tab.~\ref{tab:beta}. According
to the measured $\beta^*$ values at the minimum ($1^{st}$ row in
Tab.~\ref{tab:beta}) the beta 
function in the vertical plane exceeds the one in the horizontal plane
by 15\% in IP6 and by 25\% in IP8. In IP8 the effect would be about twice as
large as the statistical errors and should be visible in the vernier
scan data. However, as stated above, there
is no evidence for an unequality between the two planes, if anything the
vertical beam size appears to be 
smaller than the horizontal one, not larger. 
\par
\begin{table}[h!]
\renewcommand{\arraystretch}{1.0}
\begin{center}
\begin{tabular}{|l|c|c|c|c||c|c|c|c|} \hline
  & \multicolumn{4}{|c||}{STAR} &  \multicolumn{4}{c|}{PHENIX} \\ \cline{2-9}
  & BH   & BV & YH & YV & BH & BV & YH & YV \\ \hline
measured at min. & 0.64 & 0.93 & 0.75 & 0.63 & 0.76 & 0.99 & 0.61 & 0.70 \\
measured at IP & 0.65 & 1.06 & 0.82 & 0.66 & 0.82 & 1.97 & 0.63 & 0.70 \\
U12-v3         & 0.72 & 0.71 & 0.72 & 0.71 & 0.72 & 0.72 & 0.71 & 0.72  \\ \hline
\end{tabular}
\caption{\label{tab:beta}Model (U12-v3) and measured beta-functions (taken
  from~\cite{beta}) at IP6 and IP8 in units of [m].  }
\end{center}
\end{table}
Taking into account the measured 
shift of the waist of the beta function away from the center of the
interaction region ($2^{nd}$ row in Tab.~\ref{tab:beta}), the beta-function
disparity between the two planes becomes more pronounced, 20\% of the
horizontal $\beta^*$ in IP6
and 100\% in IP8. Such a large disparity as in IP8 can be excluded with
certainty. In addition to that, the measured-at-IP approach would result in a
vertical beam size in IP8 that exceeds the one in IP6 by 3 times the quoted
statistical error. As before, there is no indication for such
disparity in the data. Therefore, in this analysis, model
beta-function values were assumed. 
 
\subsection{Consistency Checks}
\begin{figure}[h!]
\begin{center}
\mbox{\epsfig{file=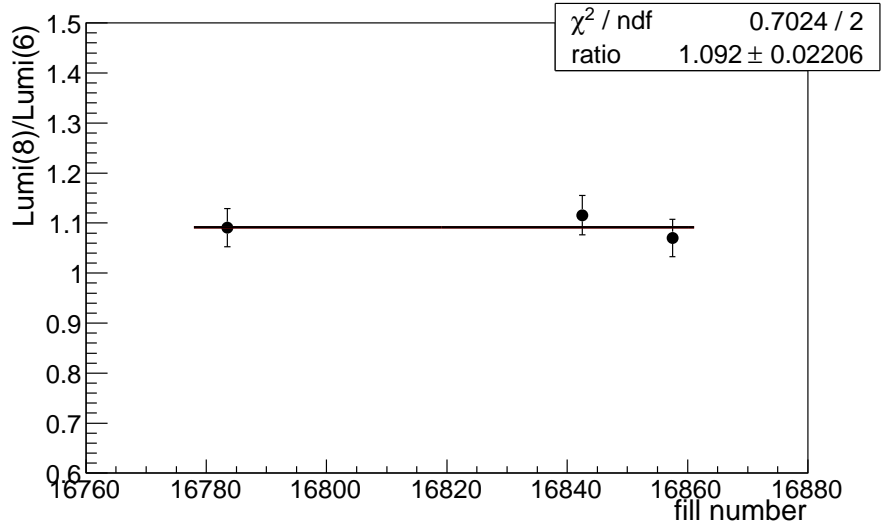,width=0.80\linewidth}}
\end{center}
\caption{\label{fig:ratio} Ratio of instantaneous luminosities in
  PHENIX and STAR
  according to Tab.~\ref{tab:xsec}. }
\end{figure}

As a consistency check, the ratio of
the instantaneous luminosities in STAR and PHENIX at the times of the
three vernier scans are shown in 
Figure~\ref{fig:ratio}. The two vernier scans in store 16783 were done
early in the store while the luminosity was still dropping more
rapidly than at the end of a store. Therefore, the instantaneous
luminosity of the later scan, the one in IP8, was scaled back to the
time of the earlier scan, the one in IP6. This was done before the
ratios were calculated. The average ratio yields 1.092. 
Due to the abort gap line-up in IP8 (compare
section~\ref{sec:fillpatt}), the ratio of the expected instantaneous 
luminosities in PHENIX and STAR corresponds to 111/102 = 1.09, the
ratio of the total number of colliding pairs. The number derived from
colliding pairs and the measured ratios are fully consistent with each
other, no other effects seem to be at work. Keep in mind that in case of significantly
different $\beta^*$-values in the two experiments the luminosity
ratio would be expected to change accordingly. There is no evidence
for unequal $\beta^*$-values. This is an additional argument to
dismiss the measured $\beta^*$-values quoted in Tab.~\ref{tab:beta}
even though the vernier scan method cannot help to determine the
individual values themselves. Vernier scans are, however, quite sensitive to a
disparity between the two IPs as well as the two planes. 
\par
The ratio of 111/102 as well as identical effective cross sections for
the two experiments are further supported by our archived data. 
\begin{figure}[h!]
\begin{center}
\mbox{\epsfig{file=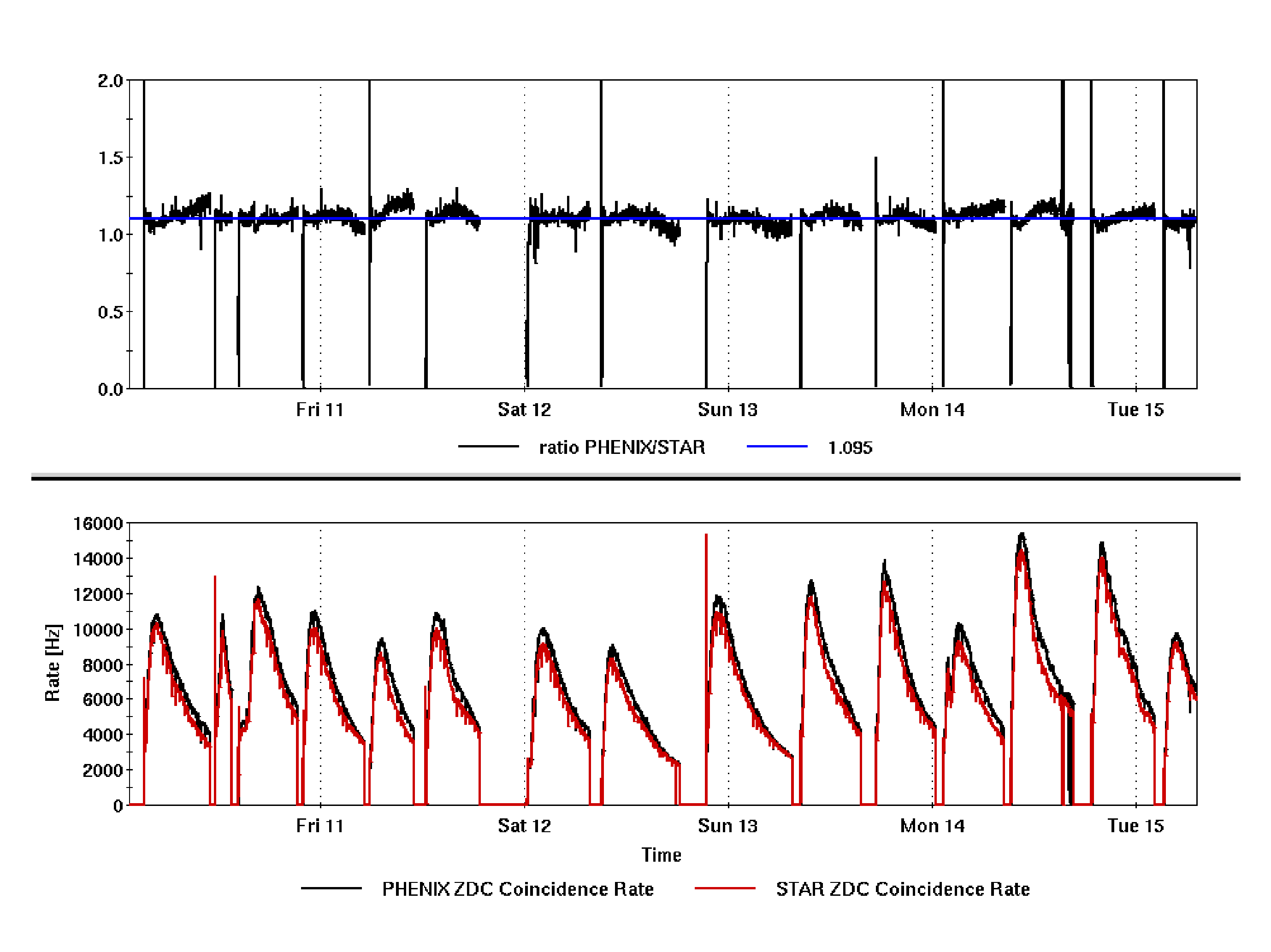,width=0.90\linewidth}}
\end{center}
\caption{\label{fig:last15} ZDC coincidence rates from the two
  experiments (bottom) and the ratio between the two (top) for the
  last 15 stores in the UU run. }
\end{figure}
Fig.~\ref{fig:last15} shows the online coincidence signals from STAR
and PHENIX (bottom) as well as the
computed ratio between the two signals (top). Data from the last 15
stores of the UU run in 2012 are included. Regular automatic
luminosity optimizations, done every 30 minutes, guaranteed optimal values for both
experiments. The fitted average ratio for those 15 stores yields a
convincing ratio of 1.095 between PHENIX and STAR luminosities. This
is in excellent agreement with the expected value of 111/102 as well as the value
from vernier scans provided model beta functions are used.

\section{\label{summary} Summary}
Combining the results, various systematic errors and the statistical
error from above, this analysis adds up to the following effective
cross sections:
\begin{itemize}
\item STAR: 15.84 b $\pm$ 1.7\% (stat.) $\pm$ 3.6\% (sys.)
\item PHENIX: 15.79 b $\pm$ 1.7\% (stat.) $\pm$ 3.6\% (sys.)
\end{itemize}
Note that the systematic error incorporates a total of 2.8\% from beam current
measurements, 1.0\% from relative beam position measurements and 2\% from
absolute beam position measurements (crossing angles). 

\section{Acknowledgements}
I thank my colleagues Christoph Montag, Mike Blaskiewicz and
Wolfram Fischer for numerous fruitful and engaging discussions on this
subject.



\begin{thebibliography}{99} 

\bibitem{VDM}
S. Van Der Meer, ISR-PO/68-31, KEK68-64.

\bibitem{hg}
M.A. Furman, M.S. Zisman ``Luminosity'', Handbook of Accelerator
Physics and Engineering, Chapter 4, P. 247-270, 1999. 

\bibitem{sc}
J.M.Brennan, M.Blaskiewicz, K. Mernick ``Stochastic Cooling in RHIC'',
IPAC12 proceedings, New Orleans, 2012. \\
M. Blaskiewicz, J. M. Brennan, K. Mernick, ``RHIC Luminosity Increase
with Bunched Beam Stochastic Cooling'',  
COOL'13 proceedings, Murren 2013. 

\bibitem{2Gauss}
K.A.Drees (BNL), S. White (CERN), ``Vernier Scan Results from the
First RHIC Proton Run at 250 GeV'', IPAC10 Proceedings, 2010.

\bibitem{zdc} A. Baltz et al., Nucl. Instr. and Methods, A417 (1998) 1

\bibitem{tnpp}
A. Drees, ``Analysis of Vernier Scans during RHIC Run-13'', C-AD AP note C-A/AP/488, 2013.  

\bibitem{ibs}
A. Piwinski, ``Touschek Effect and Intrabeam Scattering'', Handbook of
Accelerator Physics and Engineering, World 
Scientific 1999, p.125-127. 

\bibitem{dcct}
  http://www.cadops.bnl.gov/Instrumentation/InstWiki/index.php/
 RHIC\_Current\_Transformer

\bibitem{mb}
code ``lineup.f'', originally written by M. Blaskiewicz. Modified for
the purposes of this analysis in 2013. 

\bibitem{IPM}
R. Connolly et al., ``RESIDUAL-GAS-IONIZATION BEAM PROFILE MONITORS IN
RHIC'', Proceedings of BIW10 Conference, Santa Fe, New Mexico, US,
2010. 

\bibitem{beta}
M.Bai, private communication

\end{thebibliography}
\end{document}